\documentclass{kapproc} 

\RequirePackage{graphicx}%
\RequirePackage{epsf}%
\input{psfig.sty}

\upperandlowercase
\let\footnote\savefootnote
\let\footnotetext\savefootnotetext 
\let\footnoterule\savefootnoterule 
\def\degper{\ifmmode \rlap.^{\circ} \else $\rlap{.}^{\circ} $\fi}

\kluwerbib 

\begin{document}


\articletitle{The IMF of GMCs}


\chaptitlerunninghead{IMF of GMCs}



 \author{Leo Blitz, Erik Rosolowsky}
 \affil{Astronomy Department, University of California, Berkeley, CA, USA}
 \email{blitz@astro.berkeley.edu, eros@astro.berkeley.edu}





 \begin{abstract} The properties of GMCs in several Local Group
galaxies are quantified and compared.  It is found that the mass
spectrum of GMCs varies
from galaxy to galaxy.  The variations are significant and do not
appear to be the result of systematic uncertainties.  Nevertheless, it
appears that all of the GMCs follow the same size--linewidth and
mass--linewidth relations with little scatter.  The power law indices 
of these relations
imply that the GMCs are self-gravitating, and that the mean surface
density of Local Group GMCs is approximately constant.  This, in turn,
implies that the mean internal pressure of GMCs is also constant.  If
the IMF of stars is determined by a Jeans instability, this constant
internal pressure suggests that the distribution of stellar masses does not
vary significantly in galactic disks when averaged over suitably large
areas. Thus, although the distribution of GMC masses produced by
various Local Group galaxies is quite variable, the large-scale
properties of the GMCs is not.

\end{abstract}

\section{Introduction}
The IMF in galaxies is not directly measurable except in rare cases
(see Wyse this volume), but is often taken to be the same as that
determined locally (Salpeter 1955) without much justification.
Because the evolution of disk masses depends sensitively on the shape
of the IMF at the low-mass end, knowledge about the variation of the
IMF from galaxy to galaxy is an important parameter for understanding
galaxy evolution.  

One approach is to look at the GMCs from which the stars form.  If
the clouds have similar properties within and between galaxies, it
would suggest that the stars that form from the GMCs might have
similar properties and distributions.  It is only within the last few
years, however, that unbiased surveys of Local Group GMCs have been
performed at sufficient angular resolution and sensitivity to
determine the properties of GMCs in external galaxies (e.g.,
Mizuno et al.~2001a,b; Rosolowsky et al.~2003).  These surveys are
large enough that comparisons can be made with GMCs in the Milky Way
(Solomon et al.~1987; Heyer, Carpenter \& Snell 2001).  In this paper,
we look at the GMC mass functions in the Milky Way, the LMC and M33,
as well as the linewidth--size and linewidth--mass relations for these
same galaxies to see what might be inferred from the current state of
the observations.
 
\section{The mass function of GMCs}

Determination of the mass function of GMCs requires a large, unbiased
survey of the molecular gas in a galaxy at sufficient angular
resolution to separate the individual clouds from one another.  It is
not necessary to resolve the clouds in external galaxies, since the
mass is proportional to the CO flux if the CO-to-H$_2$ conversion
factor ($X$) is constant within a galaxy and from one galaxy to
another.  For the Milky Way, sufficient angular resolution has been
available since the discovery of the CO line. The problem rather had
been the large areas subtended by the CO emission from individual
clouds compared to the telescope beams and the velocity blending
produced by our edge-on view of the disk. Large surveys of the CO
emission were required to obtain reliable GMC catalogues (Dame et
al.~1986; Solomon et al.~1987).  Nevertheless, attempts to obtain an
unbiased catalogue of a sufficiently large sample of clouds suggest
that the mass function $dN/dM \propto M^{-1.6}$ (Williams \& McKee
1997).

More recently, Heyer et al.~(2001) have completed a survey of the
outer portions of the Milky Way visible from northern latitudes, and
catalogued about $10^4$ molecular clouds.  Again, resolution was not
an issue; sky coverage required more than $1.6\times 10^6$ spectra to
complete the survey.  Heyer et al.~(2001) concluded that $dN/dM
\propto M^{-1.9}$, but suggest without detailed analysis that the
power law index is not significantly different from that found by
Solomon et al.~(1987).
  
Only a few galaxies beyond the Milky Way have had complete surveys of
molecular gas done at high enough angular resolution to resolve the
emission into GMCs: M33 (Engargiola et al.~2003), the LMC (Mizuno et
al.~ 2001a), the SMC (Mizuno et al.~2001b), and IC 10 (Leroy et al.~
in preparation).  The LMC and the SMC are close enough to be mapped
with a filled aperture telescope, but each covers a large enough
fraction of the sky that a dedicated program requiring many months has
been necessary to map all of the molecular gas.  In the SMC there are
too few GMCs to obtain a reliable mass spectrum.

Beyond the Magellanic Clouds, aperture synthesis is required to
separate and resolve individual GMCs.  For a galaxy such as M33,
however, a great deal of observing time is needed to make a mosaic
large enough to cover the 0\degper5 extent of the molecular emission.
Wilson and Scoville (1990) mapped 17 fields in M33 with a 1$^{\prime}$
primary beam to obtain the first maps of individual molecular clouds
in M33.  Engargiola et al.~(2003) required almost 800 fields with a
2$^{\prime}$ beam to get a nearly complete map of the galaxy.  A
catalogue of GMCs generated from this map, superimposed on an H$\alpha$
map of the galaxy, is shown in Figure 1.  Figure 2 shows the GMCs
superimposed on the {\sc Hi} map of Deul \& van der Hulst (1987).

\begin{figure}[ht]
\centerline{\psfig{file=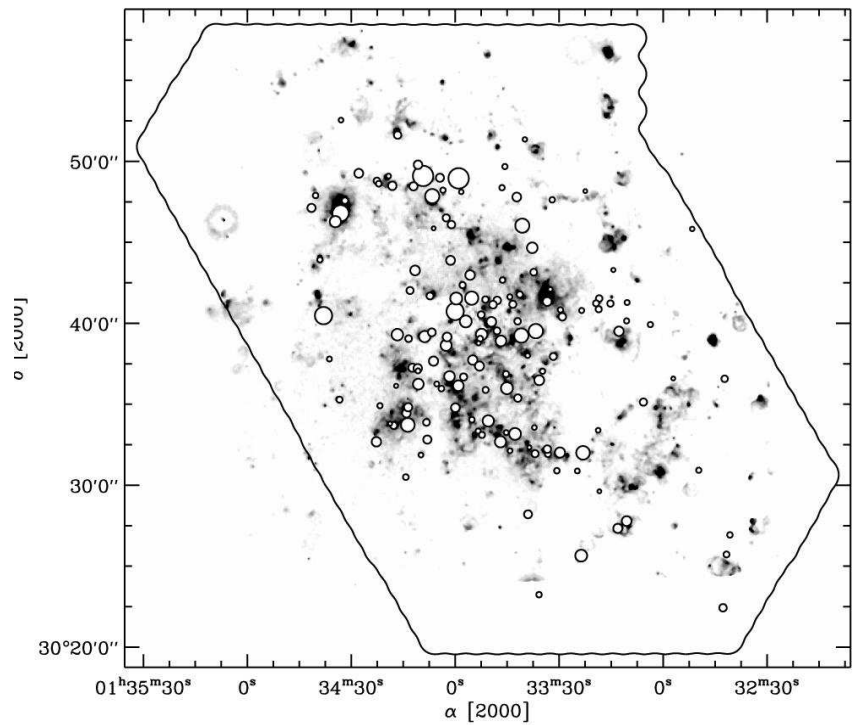,height=4in}}
\caption{The molecular clouds catalogued by Engargiola et al.~(2003),
shown at black dots enclosed by white circles, superimposed on a
continuum subtracted H$\alpha$ map from Massey et al.~(2002).  The
diameter of each dot is proportional to the H$_2$ mass of each GMC.
Note the good correspondence between the {\sc Hii} regions and the
location of the GMCs.}
\end{figure}

\begin{figure}[ht]
\centerline{\psfig{file=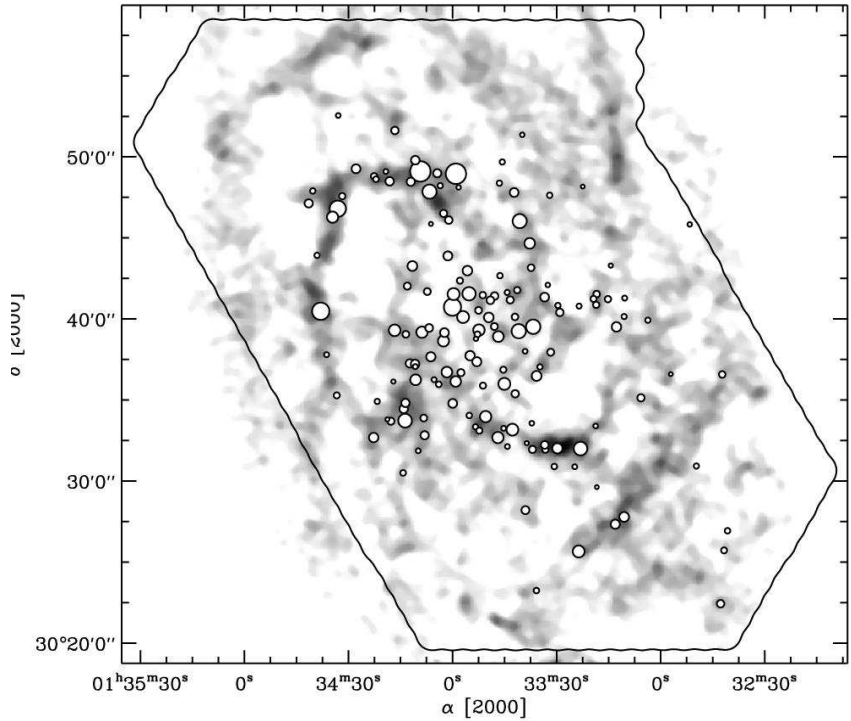,height=4in}}
\caption{The molecular clouds catalogued by Engargiola et al.~(2003)
superimposed on a map of {\sc Hi} surface density from the data of
Deul \& van der Hulst (1987). Note how the {\sc Hi} is arranged along
filaments and how the GMCs are located almost exclusively on the
filaments.}
\end{figure}

The mass spectra for M33, the LMC and the Milky Way are derived from
the GMC catalogues of Engargiola et al.~(2003), Mizuno et al.~(2001a),
Heyer et al.~(2001) and Solomon et al.~(1987) under the assumption of
a single value of $X = N(\mbox{H}_2)/I_{\mathrm{CO}} = 2 \times 10^{20} 
\mbox{cm}^{-2}~(\mbox{K km s}^{-1})^{-1}$. The total molecular mass varies
greatly among the three galaxies.  Thus, to compare the mass spectrum
of each galaxy on an equal footing, we make a plot of the cumulative
mass distributions normalized to the most massive cloud in each
galaxy.  A plot comparing the mass spectra of the the three galaxies
(showing the inner and outer Milky Way separately) is shown in Figure
3.  It can be clearly seen that the mass spectra of all three galaxies
can be well described by a power law: $dN/dM \propto M^{-\alpha}$.
The index $\alpha$ of 2.3 for M33 is significantly steeper than that
of the inner Milky Way.

\begin{figure}[ht]
\centerline{\psfig{file=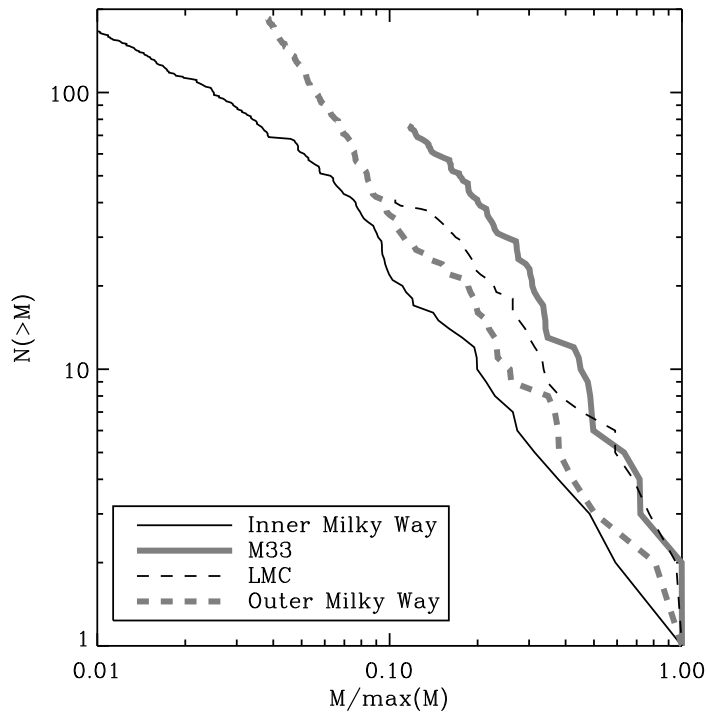,height=4in}}
\caption{The cumulative mass spectrum for the inner and outer Milky
Way, the LMC and M33.  Each mass spectrum is normalized to the most
massive cloud observed.  Note that the mass spectra for the LMC is
identical to that for M33 for the 7 most massive clouds, consistent
with the results of Mizuno (2001a).  There seems to be little question
that the mass spectrum of M33 is very different from that of the inner
Milky Way.}
\end{figure}

As a check, we ask the following question: Scaling the H$_2$ mass of
the Milky Way to that of M33, how many GMCs would one expect with a
mass greater than that of $7\times 10^5~M_{\odot}$, the largest GMC
mass in M33?  From the observations of Dame et al.~(1986) for the
largest GMCs in the Milky Way to M33, we would expect to find 15
clouds with masses larger than the largest mass GMC in M33; these 15
clouds would have a large fraction of the the total mass in GMCs in M33.
It is very unlikely that these would have been produced by the same
parent population; the difference in the mass spectra between
the inner Milky Way and M33 is apparently real. It is unclear why these
differences occur, and it is also useful to know whether the clouds
themselves have different gross properties.

The particular differences in power law index also imply fundamental
differences in the way molecular gas is distributed in the Milky Way
and in M33.  In the Milky Way, a power law index $<$ 2 implies that
most of the mass in molecular gas is in the highest mass clouds.  In
M33, the power law index is $>$ 2, which implies that most of the mass
is in the lowest mass clouds.  However, to avoid an infinite mass when
integrating the mass distribution requires either a mass cutoff or a
change in index of the mass distribution. From a knowledge of the
total molecular mass in M33, Engargiola et al.~(2003) estimate that
this change occurs at about 4--6 $\times 10^4$ M$_\odot$.  There is
thus a knee in the mass distribution and most of the GMC mass in the
galaxy occurs near the knee.  This implies that there is a
characteristic mass of the molecular clouds in M33; for some reason,
M33 primarily produces GMCs with masses of about 5 $\times~10^4$
M$_\odot$.

\section {The Properties of GMCs}

One way to compare the gross properties of individual GMCs in various
galaxies is to look at the size--linewidth relation, a comparison of
the radius of a cloud with its linewidth for many clouds.
That such a relation exists was first suggested by Larson (1981). In
the Milky Way, several investigators obtain a size--linewidth relation
for GMCs with a power law index close to 0.5 with little scatter among
the various determinations (e.g. Blitz 1993).  
This value is to be compared with that
determined in other galaxies.  Comparisons are complicated by several
factors, however.  First, most extragalactic GMCs are only marginally
resolved and it becomes necessary to deconvolve the beam from the
measurements.  Second, it is important to correct the data for
observations made with different sensitivities.  These would likely
leave the linewidths unchanged, but can give differing results for the
diameters of the clouds.  A comparison of the size--linewidth relation
is shown in Figure 4 for the Milky Way, the LMC and M33.  For the LMC,
we take the catalogue of Mizuno et al.~(2001a) and apply a correction
for the beamsize of the telescope they used.  We have only included
clouds from the outer Galaxy study of Heyer et al. (2001) that have
reliable kinematic distances and show no signs of blending.  The GMCs
in M33 and the LMC fall nicely on the relationship found for the Milky
Way.
\begin{figure}[ht]
\centerline{\psfig{file=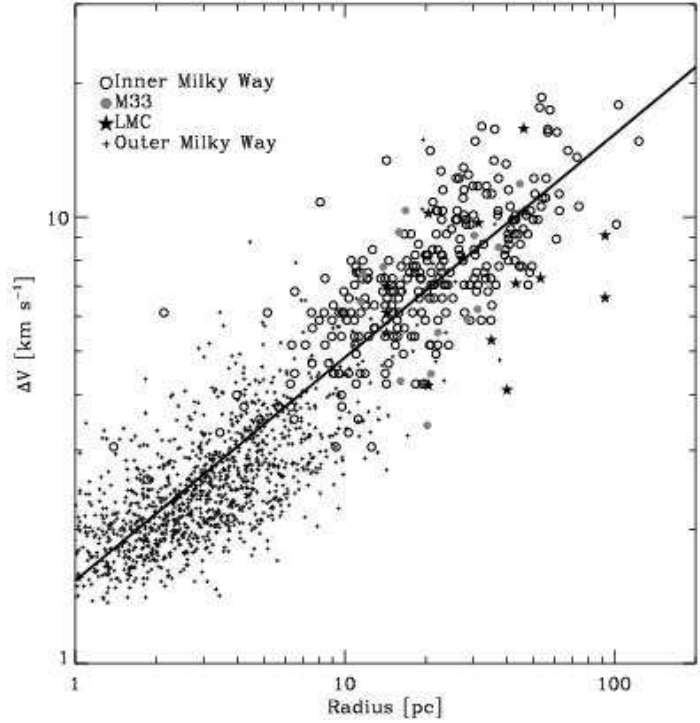,height=4in}}
\caption{The size--linewidth relation for GMCs in the galaxies
indicated.  The solid line is not a fit but rather has a power law
index of 0.5 showing that such a slope is a good fit to the data.  }
\end{figure}

One way to get around the problem of observing clouds with only a few
resolution elements across them is to look at the mass--linewidth
relation.  In this case, we plot the CO luminosity ($L_{\mathrm{CO}}$)
against linewidth ($\Delta V$) and assume that assume that
$L_{\mathrm{CO}}$ faithfully traces H$_2$ mass.  If GMCs are
self-gravitating, we have
\begin{equation}
\Delta V^2 = \alpha GM/R
\end{equation}
where $\alpha$ is a constant near unity that depends on the mass distribution.
If the clouds obey a size--linewidth relation with a power law
index of 0.5, then $\Delta V \propto R^{0.5}$.  Together, this
implies that
\begin{equation}
M/R^2 = \mbox{constant;}~~~~ M \propto \Delta V^4
\end{equation} 

\noindent The first condition implies that the surface density of GMCs
is constant, the second, that a relation for molecular clouds exists
similar to the Faber-Jackson relation for elliptical galaxies (Faber
\& Jackson 1976).  The assumption of self-gravity seems fairly
safe; The best argument for Milky Way GMCs comes from
comparisons of their internal pressures with the external pressures in
the disk that come from the hydrostatic equilibrium of the gas in the
disk.  In the solar vicinity, internal pressures are generally an
order of magnitude greater than the external pressure and the clouds
must therefore be self-gravitating if they are older than a crossing
time.  Since the properties of the catalogued extragalactic GMCs are
not grossly different from those in the Milky Way, nor are the 
values of hydrostatic pressure expected to be very different, 
this assumption seems reasonable.

Figure 4 shows a plot of $L_{\mathrm{CO}}$ vs. $\Delta V$ for the
catalogued GMCs in the Milky Way, M33 and the LMC.  The solid line is
not a fit to the data, but a line with a power law index of 4.
Clearly, this index represents the GMCs quite well; a single power law
seems to describe all of the Local Group GMCs with no offset and with
a scatter of only about a factor of two.

\begin{figure}[ht]
\centerline{\psfig{file=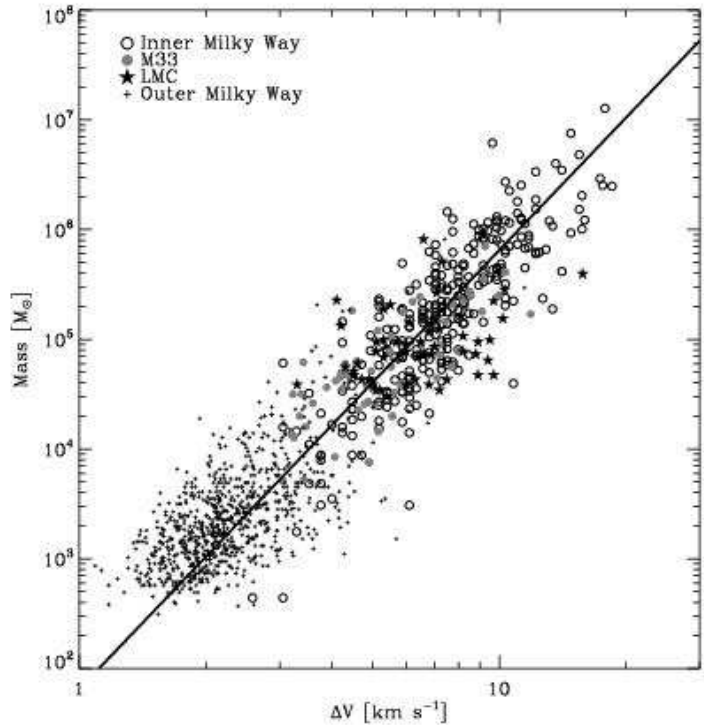,height=4in}}
\caption{The mass--linewidth relation for GMCs in the galaxies
indicated.  The solid line is not a fit, but rather has a power law
index of 4 showing that it is a good fit to the data.  A slope of 4
suggests that the clouds have a size--linewidth relation such that
$R\propto\Delta V^2$ and that the clouds are self gravitating.}
\end{figure}

Thus, the two relations suggest that the GMCs in M33 and the LMC are
similar to those in the Milky Way in that they are self-gravitating,
that they obey the same size--linewidth relation and they
have the same mean surface density with a relatively small
scatter about the mean ($\sim 100~M_{\odot} \mbox{ pc}^{-2}$). That
is, even though the GMCs in the three galaxies under consideration
here produce GMCs with different distributions of mass, the GMCs
themselves have rather similar properties.

Recall, though, that the mean internal pressure $P$ in a
self-gravitating gas cloud is proportional to the square of the
surface density $\Sigma_{gas}$ only.
\begin{equation}
P = \beta {\pi\over 2}\Sigma_{gas}^2
\end{equation}

\noindent where $\beta$ is a constant near unity depending on the
geometry of the cloud. Thus, the internal pressure of the GMCs in
these galaxies show little variation because their measured surface
densities show little variation.  
If star formation is the result of a Jeans
instability, as many astronomers believe, then the constancy of the
mean internal pressure suggests that the range of physical conditions among
star-forming GMC is quite small.  This, in turn, suggests that the
ability to form stars with mass distributions different from those
found in the Milky Way is also small. The IMF, therefore may show
little change from galaxy to galaxy within the Local Group, and very
likely, in galactic disks with properties similar to three galaxies
discussed here.


\end{document}